# SYNTHESIS AND SEARCH FOR SUPERCONDUCTIVITY IN LiBC

A. Bharathi, S. Jemima Balaselvi, M. Premila, T. N. Sairam, G. L. N. Reddy,
C. S. Sundar and Y. Hariharan

Materials Science division, IGCAR, Kalpakkam, INDIA 603102

**Abstract**

Following the recent theoretical prediction of superconductivity in hole doped LiBC by *Rosner et al* [1], we have attempted to synthesise Li deficient $Li_xBC$ ( x = 1, 0.8, 0.6 and 0.4) and look for superconductivity in this system. Our synthesis procedure, following the recipe for $MgB_2$, involves reaction of elemental components in a Ta crucible at $900^o$ C under 50 bar of argon pressure. X-ray diffraction measurements indicate the formation of $P6_3/mmc$ structure up to x=0.6. However, no diamagnetic signal or zero resistance, corresponding to the superconducting transition, was observed in the temperature range of 300 to 4 K. This is possibly related to the presence of disorder in the B-C stacking; evidence for which is suggested from a study of the vibrational modes of $Li_xBC$ through infrared spectroscopy.

## 1. INTRODUCTION

Since the discovery of superconductivity in $MgB_2$[2,3] there has been a serious effort to unravel the mechanism of superconductivity in this compound[4]. This high $T_c$ in $MgB_2$, is thought to arise due to strong electron-phonon coupling of the holes in the σ band with the bond stretching $E_{2g}$ modes in the B planes[5]. Electronic band structure calculations[5] also show that the hole density of states is 2-dimensional in character, the energy dependence is flat below $E_F$, but falls off above suggesting that the density of states at $E_F$, $N(E_F)$, is not expected to change with hole doping, whereas it will decrease with electron doping. Attempts to increase $T_c$ in $MgB_2$ by chemical substitution have not thus far been successful. $T_c$ is observed to decrease by electron[6,7] doping, as seen in Al and C substitution experiments. By the substitution of Li, which corresponds to hole doping[8] $T_c$ remains constant. Whereas by Be substitution, which also results in hole doping,[9] $T_c$ shows a precipitous decrease,[10] which is understood in terms a decrease in the electron phonon coupling strength.[11] It therefore appears that the route to increase $T_c$ is to increase the electron-phonon coupling strength with the hole concentration remaining fixed atleast at the $MgB_2$ value. Along these lines[11] it was suggested that the

$CuB_{2-x}C_x$ system would be a likely candidate for high $T_c$, with Cu substitution resulting in the right magnitude of $N(E_F)$, but with an increased electron-phonon coupling due to the C substitution of the B layers. Experimental verification of these ideas has been hampered by the lack of solubility of Cu in $MgB_2$ and the fact that the $CuB_{2-x}C_x$ compound does not form. The $T_c$ in other iso-structural di-borides viz., $ZrB_2$[12], $TaB_2$[13] and $MoB_2$ stabilized with 4% $Zr$[14] have been disappointingly low. This is rationalized, based on their electronic structure,[15,16] to arise because the 4d electrons of the transition metals span the Fermi level and that the σ bonding electrons of B are ~2-4 eV below $E_F$.

Recently band structure calculations[1], have suggested a possibility of the occurrence of superconductivity in an isostructural compound LiBC. The electron count in LiBC is identical to that in $MgB_2$ with Li providing one less electron than Mg, and the replacement of B with C, providing the extra electron. Stoichiometric LiBC is a large gap semiconductor,[1,17] and it has been reported[1,17] that $Li_xBC$ can be synthesized upto x=0.24. These band structure calculations predict that by introducing Li off-stoichiometry a finite density of states appears at the Fermi level, rendering the system metallic. Further the calculation of the phonon dispersion relations in the off-stoichiometric $Li_{0.75}BC$ show that the Raman active, $E_{2g}$ mode dramatically downshifts acquiring a large line width, indicating the presence of a very strong electron-phonon coupling in this sytem.[18] It is argued that this large electron-phonon coupling could result in the Li deficient compound becoming a superconductor, in which $T_c$ can be varied by varying the extent of hole concentration, and that the $T_c$ can be as high as 115 K for a Li stoichiometry of 0.5.

Given these exciting predictions, we have carried out experiments on the synthesis of $Li_xBC$ samples for x in the range of 1.0 to 0.40, by solid-vapour phase reaction similar to that employed in the synthesis of $MgB_2$[7]. The samples have been characterised for their phase purity by X-ray diffraction. The phonon modes have been measured using IR spectroscopy in both the transmission and reflection geometries. AC susceptibility and DC resistivity has been carried out on all the samples in the 4 K to 300 K temperature range to scan for superconductivity. In contrast to the theoretical predictions, no signature of superconductivity was observed in our experiments. IR measurements suggest the presence of disorder in the B-C stacking which could account for the absence of superconductivity in these samples.

## 2. EXPERIMENTAL DETAILS

The $Li_xBC$ samples were prepared from stoichiometric quantities of the Li (99%), amorphous boron (97%) and C obtained from fullerene soot. The B and C were initially ground together and pelletised and loaded into Ta crucibles. Li was weighed in a dry box and loaded into the Ta crucibles along with the boron-carbon pellets. The crucibles were then transferred into an SS tube into which 50 bar of argon could be pressure sealed with the help of a high pressure valve. This SS tubing assembly was placed in a vacuum jacket, which was under a dynamic vacuum of $10^{-5}$ torr. The initial mixture was heat treated in a tubular furnace for 1.5 hours at 900°C. X-ray diffraction of the sample taken at this point was clearly multiphasic, eventhough the presence of a sizeable fraction of LiBC phase could be clearly discerned. The heat treated powder was seen to be inhomogeneous even under visual examination. At this stage the sample was ground thoroughly in a dry box after which it was transferred to a Ta crucible and heat treated at 900°C for a further period of 1.5 hours under 50 bar of argon. After the second heat treatment the sample was pelletised in vacuum and sintered for 4 hours. No perceptible weight loss was seen in the samples even after the above three heat treatments, suggesting the preservation of the starting stoichiometry in all samples. The retrieved pellet was 80% dense. It was noticed that the samples were extremely air sensitive, i.e., the surface visibly deteriorated under ambient conditions and had to be handled always in a dry environment. AC susceptibility measurements were carried out in a home built mutual inductance bridge apparatus operating at 941Hz. The samples were loaded into the susceptibility capsules in the dry box. The resistivity contacts were made with silver paste under an IR lamp very quickly to avoid contamination in air. The resistivity measurements were done in the four probe geometry at 30mA current. The temperature variation from 4 K to 300 K was achieved in a dipstick arrangement. The resistance and temperature data were logged online onto a personal computer through an IEEE interface card. The XRD measurements were done using Cu $K_\alpha$ radiation, in a STOE diffractometer in a Bragg-Brentano geometry, with samples loaded into 0.3 mm capillary tubes and sealed with wax in the dry box. The IR spectroscopy measurements were carried out at room temperature in a BOMEM DA-8 FTIR spectrometer in the 125 $cm^{-1}$ and 4000 $cm^{-1}$ wavenumber range at a resolution of 4 $cm^{-1}$.

## 3. RESULTS AND DISCUSSION

### 3.1 Structure

The XRD pattern for stoichiometric LiBC after the heat treatment for 3 hours with one intermediate grinding at 900$^o$C is shown in Fig.1a. It can be seen that the lines due to LiBC are all present. It is also seen that the line positions are in good agreement with that reported for LiBC in the JCPDS data file (85-2010) (cf. Fig. 1a), shown as dotted lines in the figure. The variation of the XRD pattern with further sintering treatment at 900$^o$C is shown in panel (b) and (c) of Fig. 1. It can be seen from the figure that the intensity ratio of the (102)/(002) peaks increases with the time duration of the heat treatment, in better agreement with the JCPDS data. Some impurity peaks were seen in the data that could not be indexed, but are not large in intensity indicating that the sample is predominantly monophasic LiBC. The impurity peaks also pick up in intensity with sintering treatment, possibly because of the unavoidable exposure to air during pelletisation. The XRD patterns of LiBC for the various Li contents studied are shown in Fig.2. All samples have been given the initial 3 hours treatment at 900$^o$C, with intermediate grinding followed by pelletising and heat treatment for 4 hours. It is apparent from Fig.2 that while the LiBC phase has formed in the x=1.00 and x=0.80 samples, in the samples with x=0.60 and x=0.40, a significant amorphous background is present. The peak positions (002) and (102) are seen to shift systematically with Li off stoichiometry in the three samples studied (cf. Fig.2). The lattice parameters were computed for the samples with x=1.0 and x=0.8 using the STOE program. The lattice parameter value in x=1.0 sample is a=2.7462(9)A$^o$, c=2x3.5390(10)A$^o$ with an unit cell volume of 2x23.114(11)(A$^o$)$^3$ and that in the x=0.8 sample was a=2.7396(21)A$^o$, c=2x3.5351(13)A$^o$ with a volume of 2x22.978(15) )(A$^o$)$^3$. The volume is seen to decrease with Li deficiency. The lattice parameter could not be evaluated for the samples with lower Li stoichiometry as only two peaks could be discerned in the XRD patterns (cf. Fig.2).

### 3.2 Scan for superconductivity and resistivity studies

Preliminary search for superconductivity in the off-stoichiometric $Li_xBC$ samples were carried out in the 4-300 K temperature range by the AC susceptibility technique. No diamagnetic signal could be detected in any of the samples studied, within the detectable limit of 2-3% volume fraction. The resistivity measured in the 4-300 K range in the five samples are shown in Fig. 3. The room temperature resistivity in LiBC is 0.1405 $\Omega$-cm as measured in the van der Pauw geometry. In the $Li_{0.8}BC$ sample the resistivity increases to 0.328 $\Omega$-cm. For lower concentrations the room temperature resistivity further increases. It is clear from the figure that in LiBC the d$\rho$/dT is negative. The resistivity versus temperature fits well to the variable range transport mechanism,[19] $\rho = \rho_o \exp(-(T_0/T)^d)$ with an exponent d of 1/4 and does not fit to an Arrhenius behaviour in the entire temperature range. The fit to the variable range hopping (VRH) behaviour is found to be very good in the 170-270 K region. The fitted and the measured curves are compared in the inset of Fig.3 for LiBC. Below 170 K the resistivity is seen to deviate from VRH behaviour very significantly. The resistivity curves in the Li deficient samples shown in Fig. 3 are seen to be similar to that in LiBC and also fit to the VRH behaviour in the 170 K and 270 K temperature range rather than to the Arhenius behaviour and deviate from VRH behaviour for temperatures less than 170 K. However no systematics in the variation of hopping parameter $T_0$ with Li deficiency could be discerned.

In addition to studies on $Li_xBC$ samples we have also synthesized $Li_{0.8}Mg_{0.2}B_2$. The XRD pattern for Mg substituted samples showed apart from the presence of the LiBC phase the presence of MgO of ~7.5%. The resistivity in the Mg substituted sample at room temperature is the lowest and is 0.0667 $\Omega$-cm. The temperature dependence of the resisitivity plotted for this in the same scale as the LiBC samples and is shown in Fig. 3. The resistivity is seen to be temperature independent. However, plotted in an enlarged scale this has also a negative d$\rho$/dT, and could also be fitted to the VRH rather than to an Arrhenius behaviour.

Since the resistivity data fit to VRH transport in LiBC, it appears that the electrons are trapped at in regions of disorder and result in localized states. It therefore appears that the present LiBC sample is not a band semiconductor as foreseen in the band structure calculations.[1] Hole doping in such a system does not seem to have been achieved as seen from an increase in $\rho$ with Li doping. It is pertinent to remark at this point that the calculations[1] also emphasise the importance of presence of B-C ordering, (AB stacking) on the semiconducting behaviour in LiBC. AA stacking of the layers, which introduce B-B and C-C bonds along the c-axis, result in the compound becoming semimetallic. The small increase in the resistivity with decrease in temperature (in comparison to that expected for a 1.0 eV semiconductor gap) observed in the present measurement on the LiBC sample could also possibly arise on account of pockets of semimetallic LiBC present due to B-C disorder, evidence for which is brought out from IR measurements.

## 3.3 Infrared spectroscopy studies

To obtain experimental information on the phonon modes and their variation with Li deficiency, we have carried out infrared absorption and reflectivity measurements. Factor group analysis[20] predicts 15 zone center modes for the LiBC system : $2 A_{2u} + B_{2u} + 2B_{1g} + 2 E_{1u} + 2 E_{2g} + E_{2u}$, of which the $2A_{2u}$ and $2 E_{1u}$ modes are infrared active and the $2E_{2g}$ modes are Raman active. Theoretical calculations[18] indicate that in LiBC, these 15 zone center modes span a range upto 160 meV. In the metallic $Li_{.75}BC$ system, while most of the modes are relatively unaffected, the Raman active $E_{2g}$ mode at ~ 150 meV shows a dramatic down shift to ~ 85 meV and acquires a width comparable to the mode frequency. This mode is shown[18] to dominate the electron-phonon interaction responsible for the suggested superconductivity in $Li_{.75}BC$.

The results of our infrared absorption measurements are shown in Fig.4. The absorption spectrum in LiBC is characterized by several sharp features in the range up to 800 cm$^{-1}$, and isolated strong absorption features centered at 980 cm$^{-1}$ 1180 cm$^{-1}$ and 1330 cm$^{-1}$. At first it can be seen that the absorption spectrum bears a strong resemblance to the calculated phonon density of states,[18] shown in the top panel rather than the expected 4 IR modes. This suggests the presence of disorder even in the stoichiometric LiBC, leading to the relaxation of the selection rules. The feature at 980 cm$^{-1}$ is characteristic of B-B vibration as seen in amorphous B[21] and the mode at 1330 cm$^{-1}$ is similar to the disorder mode in graphite.[22] While the presence of these features can apriori be attributed to the presence of unreacted B and C, the fact that these mode frequencies change systematically with Li deficiency (see below) suggests that these are related to B-B and C-C vibrations in the LiBC phase, which can arise due to B-C disorder, both in-plane and along the c-direction.

In addition to IR transmission measurements in a KBr matrix, we have also carried out reflection measurements on sintered polished pellets of LiBC and obtained the absorption spectrum through a Kramers-Kronig analysis. The resulting absorption curve is shown in the top panel of Fig.5. It is seen that there is a strong similarity with the absorption obtained from direct measurements (bottom panel) and further the results from reflectivity measurements extend the range below 280 cm$^{-1}$, a restriction imposed in the direct absorption measurements due to absorption in the KBr matrix.

These absorption curves as shown in Fig. 5 have been analysed in terms of sum of Lorentzians each of which represents a spectral feature in the absorption curve. The resulting "mode" frequencies are summarized in table 1, along with the results of theoretical calculations[18]. This helps to provide a tentative assignment of the features seen in the absorption spectrum in terms of zone center modes. We note that the calculated features[18] corresponding to both LiBC and Li$_{.75}$BC shows up in the measured absorption spectrum of LiBC. It is interesting to note that the theoretically calculated[18] Raman modes at 1180 and 1194 cm$^{-1}$ shows up as a dominant feature in the infrared absorption spectra – again indicating the presence of disorder.

The variation of 'mode' frequencies with Li content is summarized in Fig.6. It is noted that several of the modes, including the 1330 cm$^{-1}$ mode, hardens with Li deficiency indicating that it arises from the disordered LiBC phase rather than from an impurity phase of unreacted amorphous carbon. The only notable exception is the feature at 1100 cm$^{-1}$ that shows a distinct softening. The hardening of modes may be linked to the reduction of cell parameters, as indicated by XRD measurements. One of the definitive and important findings of the theoretical calculations[18] is the dramatic softening of the Raman mode with Li deficiency. This is not seen in the present experiments (cf. Fig. 4), indicating that hole doping has not been achieved in our Li deficient samples. This may be related to the disorder of B-C stacking.

## 4. SUMMARY AND CONCLUSIONS

Li$_x$BC samples have been prepared under a high pressure of Ar by solid vapour reaction, similar to MgB$_2$. The samples were characterized by XRD and IR spectroscopy. It is seen that the phase forms for Li concentration of x=0.60, the bulk of the x=0.40 sample is possibly amorphous, although IR measurements imply the presence of short range order in this sample. The lattice volume shrinks with

Li deficiency and this is consistent with some of the IR modes hardening with decrease in Li concentration. The absence of superconductivity in the system in all the samples with Li deficiency in the 4-300 K range is in distinct contrast to theoretical predictions based on phonon mediated mechanism of superconductivity.[1] IR spectroscopy reveals that none of the modes show a distinct softening nor broadening, with Li deficiency, also in contrast to theoretical predictions[18]. Temperature dependence of resistivity in LiBC is typical of that observed in a disordered system. The absence of superconductivity in our samples could possibly arise on account B-C disorder, as is brought out from the IR measurements, argued to be crucial for the electronic structure in stochiometric LiBC.[1] In order to bring about the B-C ordering, annealing treatments at low temperature are underway. Preliminary measurements reveal that annealing LiBC at $400^0$C for 40 hours under 50 bar of Ar results in a 50-fold increase in the room temperature resistivity, characterization by IR is in progress. Unless the presence of B-C order in the LiBC sample is ensured, the validity of the theoretical predictions for this system[1,18] cannot be ascertained.

Our resistivity curve in polycrystalline LiBC (cf. Fig. 3) is very similar to that obtained recently on single crystals of LiBC.[23] The mode at 1180 cm$^{-1}$ seen in the reflectivity spectrum reported on their samples is also seen in our samples, however, it is noteworthy that our synthesis procedure is entirely different from that adopted earlier.[17,23]

## ACKNOWLEDGMENTS

The authors sincerely thank Dr. J.M. An and Prof. W.E. Pickett for providing us their unpublished results on zone center phonon frequencies in LiBC and Li$_{.75}$BC that has helped in the interpretation of infrared absorption results.

# FIGURE CAPTIONS

Fig.1 XRD patterns in LiBC, (a) Treated at 900$^o$C for a total of 3 hours with one intermediate grinding (b) same as (a) and sintered at 900$^o$C for 1 hour (c) same (a) sintered at 900$^o$C for 4 hours. Dotted lines are line positions from JCPDS file.

Fig.2 XRD patterns of Li$_x$BC, for (a) x=1.0, (b) x=0.8, (c) x=0.6 and (d) x=0.4

Fig. 3 Resistivity as a function of temperature in Li$_x$BC samples and in Li$_{0.8}$Mg$_{0.2}$BC samples, the rsesitivity axis for Li$_{0.4}$BC is shown on the right. Inset shows the fit of resistivity in LiBC to the VRH formula (solid line), open circle are the data points.

Fig.4. Infrared absorption spectra in (a) LiBC (b) Li$_{.8}$BC (c) Li$_{.6}$BC and (d) Li$_{.4}$BC. In these curves a linear background has been subtracted . The top panel shows the phonon density of states from Ref. [18].

Fig.5. Absorbance in LiBC as obtained from (a) Reflectivity and (b) Transmission measurements. These have been fitted to a sum of Lorentzians as indicated. The positions of the calculated[18] mode frequencies in LiBC and Li$_{.75}$BC are indicated by continuous and dashed sticks

Fig.6. Variation of "mode" frequencies as a function of Li deficiency. The straight lines are guide to the eye. While most of the modes harden with Li deficiency, a softening is seen for the 1100 cm$^{-1}$ mode.

Table I. Vibrational mode frequencies in cm$^{-1}$ in Li$_x$BC as obtained from the decomposition of infrared absorption in terms of Lorentzians. The numbers in the parenthesis indicate the corresponding widths. The table also indicates the theoretically calculated[18] zone center mode frequencies and their multiplicities in parenthesis.

| LiBC | Li$_{0.8}$BC | Li$_{0.6}$BC | Li$_{0.4}$BC | LiBC Theor. | Li$_{0.75}$BC Theor. |
|---|---|---|---|---|---|
| 201(17) | 212(24) | | | 169(2) | 177(2) |
| 234(42) | 239(18) | | | | |
| 288(63) | 277(62) | | | 286 | 269 |
| | | | | 303(2) | |
| 337(19) | 338(16) | 334(27) | 332(45) | | 331(2) |
| 362(27) | 363(23) | 362(30) | 362(33) | 349(2) | |
| 389(35) | 387(35) | 390(32) | 391(39) | | 373(2) |
| 422(44) | 430(53) | 428(54) | 430(54) | 418 | |
| 460(51) | 467(50) | 467(50) | 471(43) | | 454 |
| 503(53) | 498(58) | 500(59) | 503(57) | | 498 |
| 544(51) | 542(58) | 546(49) | 552(66) | 535 | |
| 594(47) | 596(67) | 593(85) | 595(99) | | 675(2) |
| 696(48) | 692(35) | 690(47) | 704(60) | | 687(2) |
| 754(30) | | | | | |
| 794(28) | | | | 795 | 805 |
| 815(12) | 814(22) | 826(10) | 840(11) | 815 | 821 |
| 1105(25) | 1100(15) | 1102(23) | 1087(17) | | |
| 1180(69) | 1188(61) | 1187(59) | 1192(58) | 1185(2) | |
| 1211(59) | 1214(54) | 1213(49) | 1218(45) | 1194(2) | |
| 1254(26) | 1253(31) | 1252(25) | 1253(23) | | |
| 1314(15) | 1323(19) | 1326(17) | 1330(12) | | |
| 1330(13) | 1336(12) | 1338(13) | 1340(12) | | |

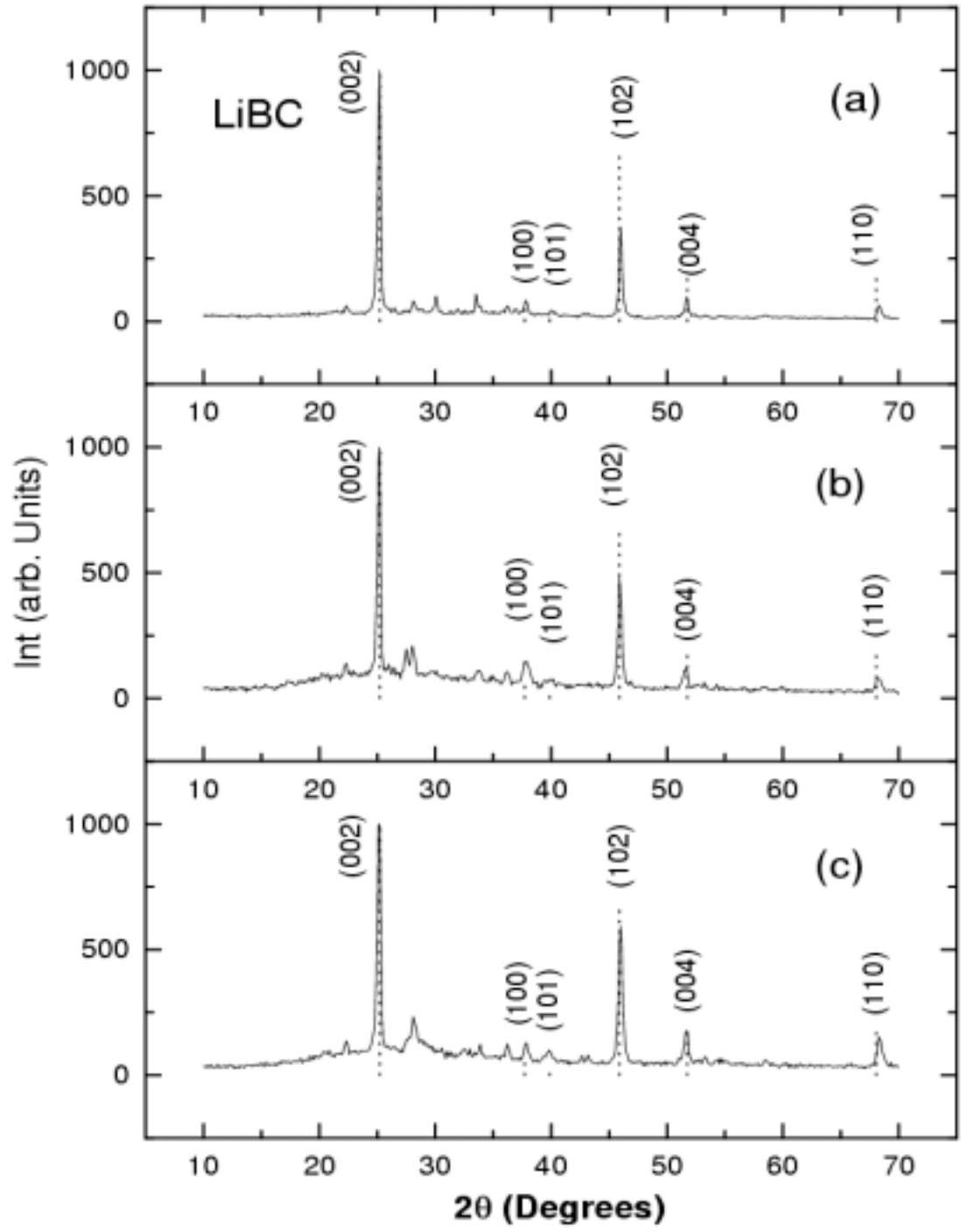

Figure.1 Bharathi et. al

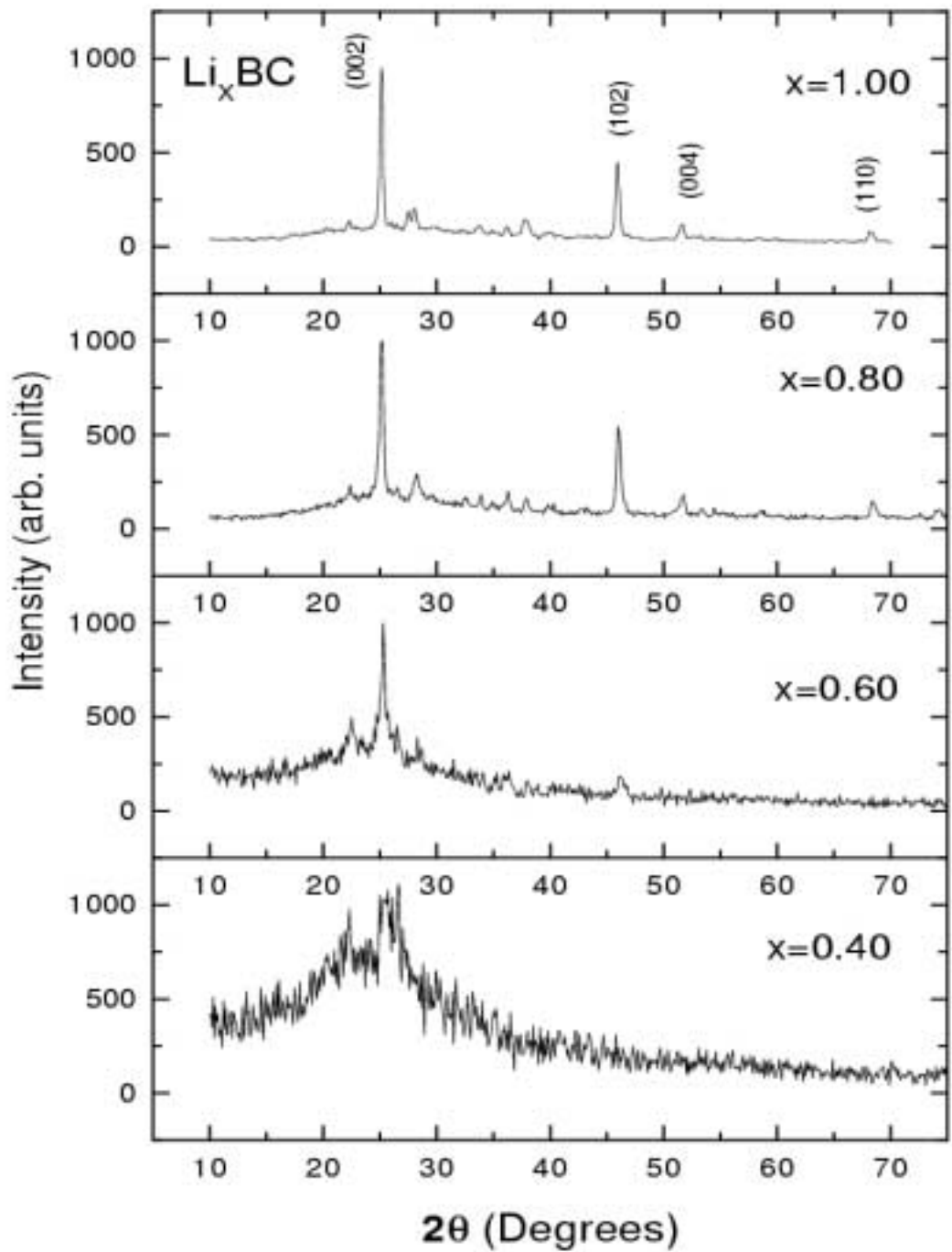

Figure. 2 Bharathi et. al

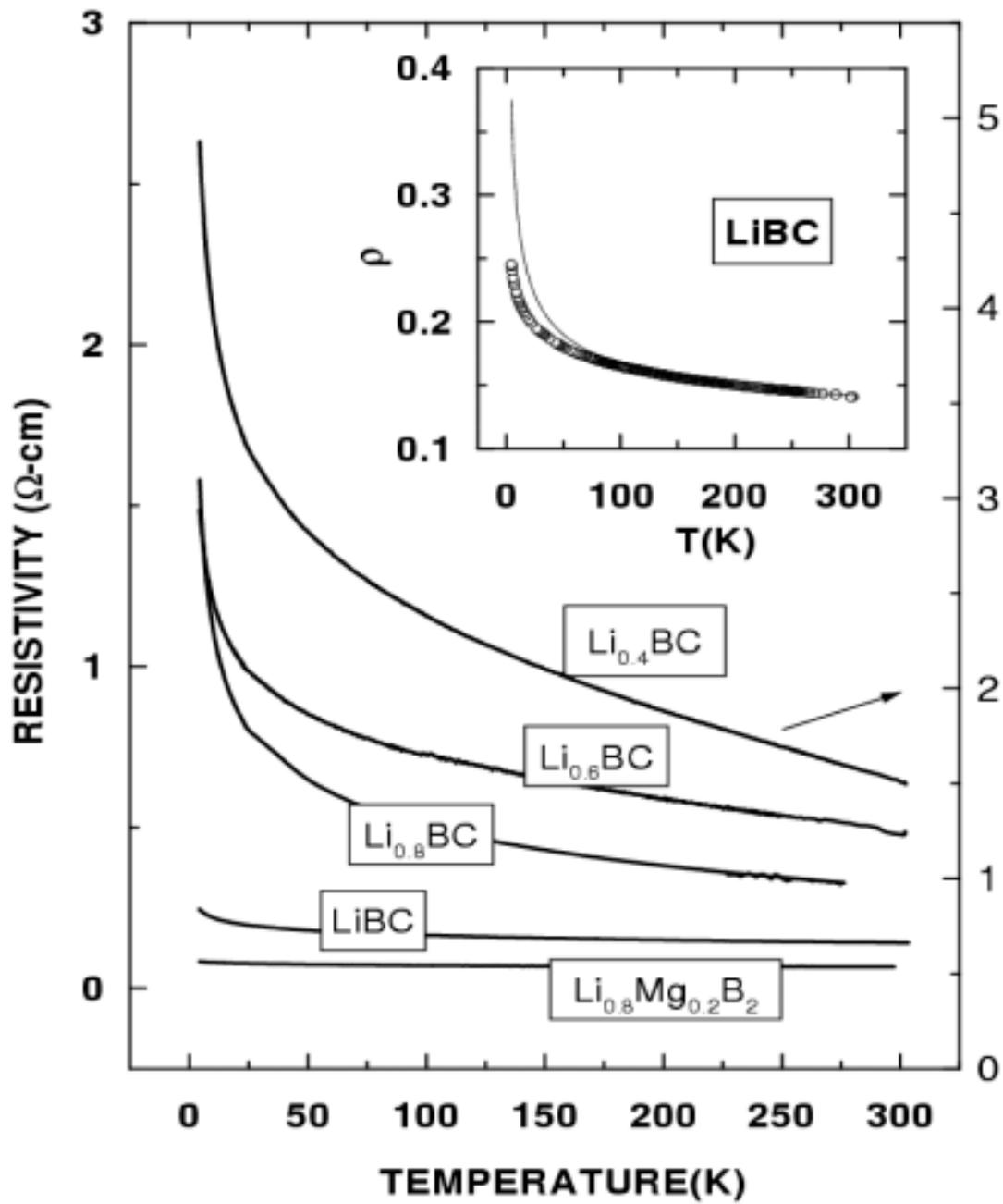

Figure. 3 Bharathi et. al

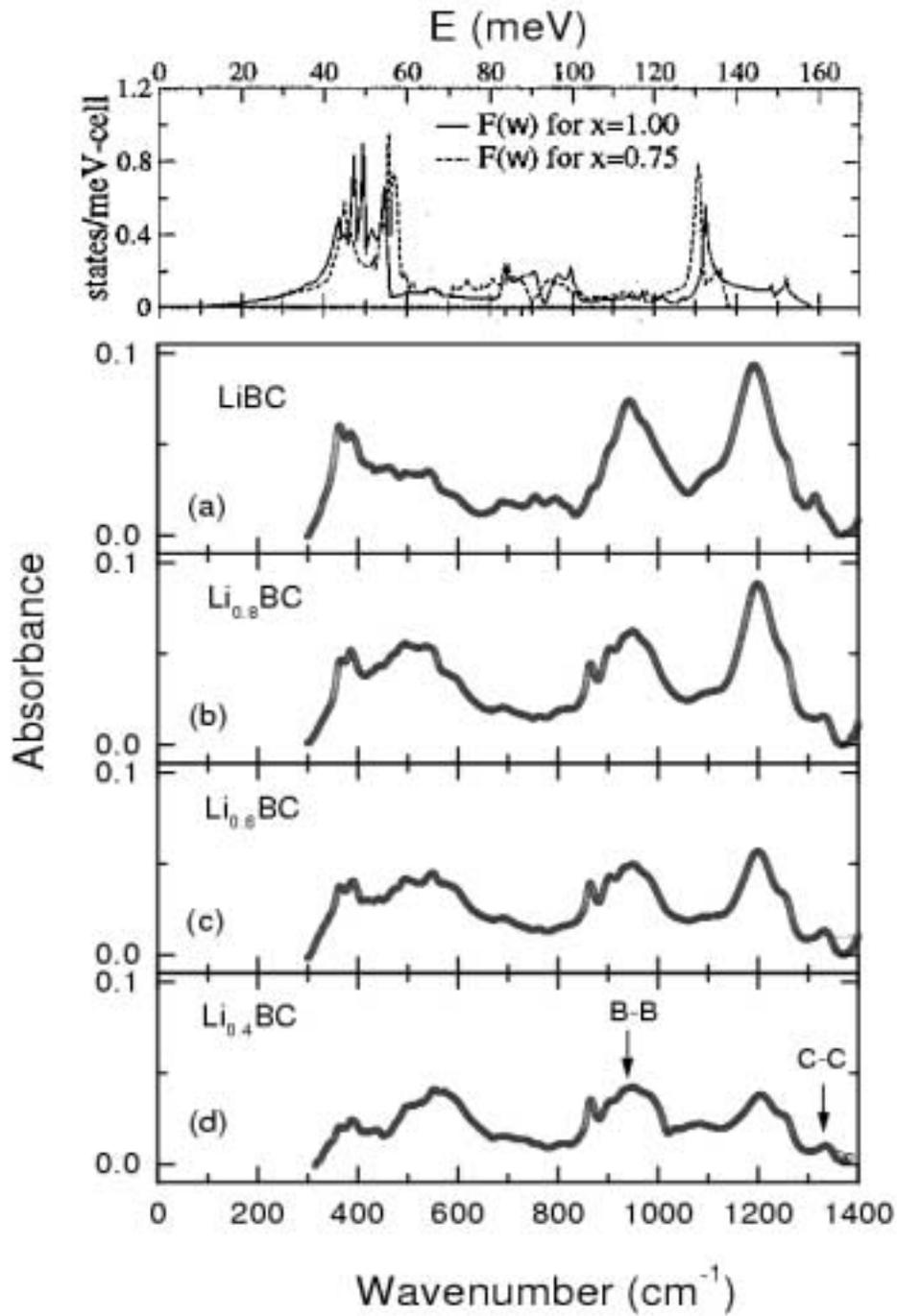

Figure. 4 Bharathi et. al

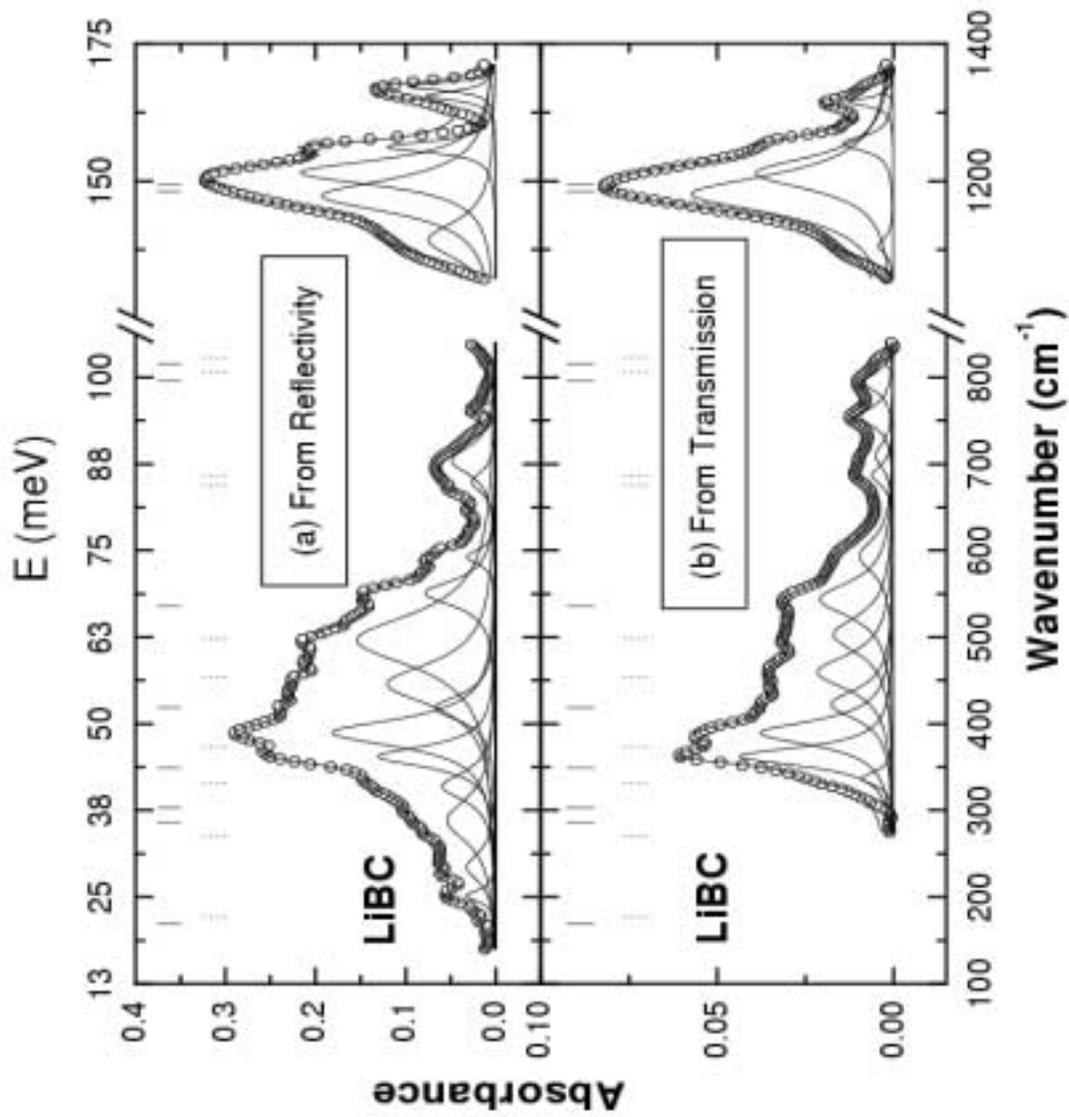

Figure. 5 Bharathi et. al

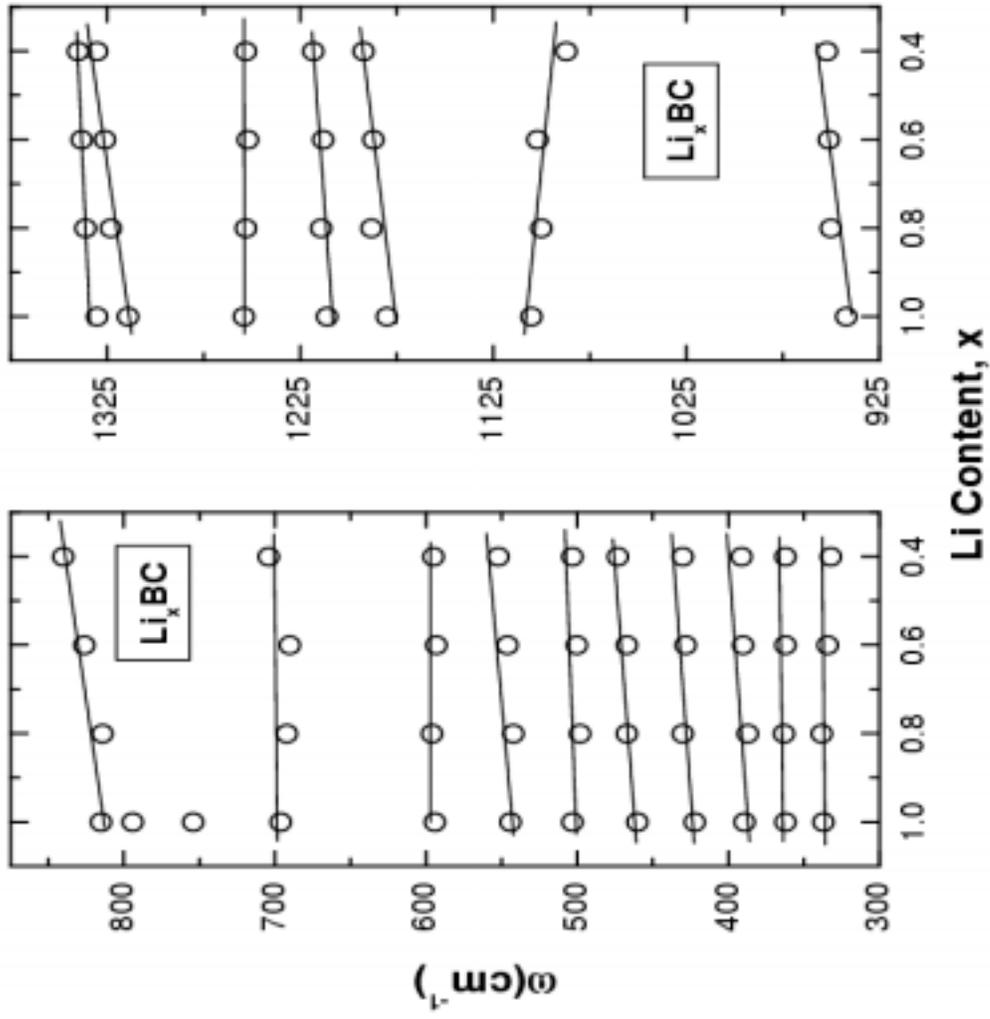

Figure. 6 Bharathi et. al